# Development of a RFID sensitive tag dedicated to the monitoring of the environmental corrosiveness for indoor applications

I. El Masri[1], B. Lescop[1], P. Talbot[1], G. Nguyen Vien[1], J. Becker[2], D. Thierry[2], S. Rioual[1]

[1]Univ Brest, Lab-STICC, CNRS, UMR 6285, F-29200 Brest, France

[2]Institut de la Corrosion, 220 rue Pierre Rivoalon, 29200 Brest, France

**Abstract.** The environmental corrosiveness is governed for indoor applications by the presence of gaseous pollutants in air and levels of temperature and relative humidity. Its determination is a challenging task and requires the monitoring of thickness reduction of selected metals in the range of few tens of nanometers. The present work aims at developing an UHF RFID sensor dedicated to such measurements. The sensor is based on the coupling between the antenna of a commercial RFID tag and a thin layer of copper exposed to the environment. The ability of the proposed sensor to be sensitive to a variation of the metal thickness in the range of tens of nanometers is demonstrated experimentally through exposure tests in a climatic chamber. The results are supported by electromagnetic simulations performed in the case of a coupling between a dipolar antenna and a thin metallic layer.

# 1. Introduction

Atmospheric corrosion of metals is a process which affects most of the metallic objects everywhere in the world due to their chemical interaction with the environment. The number of impacted industrial applications and the associated financial costs are therefore impressive. For indoor environments, the corrosivity of metals is mainly governed by the presence of gaseous pollutants and levels of temperature/relative humidity. It can be categorized on the grounds of corrosion rate *i.e.* the loss of metal by time unit. For indoor applications, the ISO 11844-1standard [1] provides the corrosion rate and IC corrosivity classes which range from IC1 (very low corrosivity) to IC5 (very high corrosivity). The pollutants may be of diverse nature (organic acids, $H_2S$, $NH_3$ …) and depend on the considered application and/or localization. Steel, zinc, copper and silver are sensitive to different pollutants and are therefore selected in the ISO 11844-1standard [1].

During the last decades, electrical resistance (ER) sensors have been developed for the determination of the IC corrosivity classes. In particular, the AirCorr$^{TM}$ sensor was commercialized within a project supported by the European Commission (Protection of cultural heritage by real-time corrosion monitoring) and dedicated to the monitoring of corrosive atmospheres in the cultural heritage sector [2-4]. In this case, the sensitive part the sensor is constituted by a strip elaborated in copper or silver due to the sensitivity of such metals to different pollutants. From the change in geometry of this metallic element (*i.e.,* thinning) due to corrosion, it is possible to calculate the loss of metal thickness and hence the corrosivity category IC. The main improvement of the method associated with the AirCorr$^{TM}$ sensor concerns its ability to detect very low metal losses at the nanometer scale. This was achieved by producing sensitive thin films by Physical Vapor Deposition (PVD) and compensating temperature variations. Detection of very low corrosion rate, typically 0.015 nm/day, authorizes the monitoring of corrosion for the very low corrosivity class IC1.

More recently, several passive RFID sensors sensitive to corrosion or cracks formation were proposed [5]. Among them, sensors based on the UHF RFID technology operating at 868 MHz or 2.45 GHz are particularly attractive. Indeed, with this technology, it is possible to develop flexible passive sensitive tags with low dimensions at very low prices and to extend largely the reading distance with respect to low-frequency RFID technology. For this purpose, several architectures can be selected as detailed in several reviews [6-10]. We recently proposed a new technique for atmospheric corrosion monitoring based on RFID chipless technology in the UHF frequency range. The variation of propagation of electromagnetic (EM) waves in a sensitive element similar to that used for ER sensors and produced by PVD was considered. Promising results were obtained for zinc transmission line and resonators during their corrosion [11-14]. The critical thickness and thus the sensitivity of the method is strongly related to the skin depth which is about few micrometers in the UHF frequency range. As the consequence, determining IC classes for indoor applications which requires the detection of thickness losses in the range of tens of nanometers is not feasible.

Another strategy was considered by several authors for the detection of corrosion of steel. Zhang *et al.* [15] proposed the design of an UHF RFID tag to probe corrosion phenomena by attaching the tag to the steel surface. The method is based on the coupling between the antenna and the underlying material which is changing from steel to corrosion products. A variation of the antenna's impedance leads to the detection of corrosion products with respect to steel. Recently, a similar approach was considered by Soodmand *et al.* [16]. A shift of the resonance frequency of a high quality factor antenna was proved to be an indicator for the detection of corrosion of steel. It is ascribed to the formation of corrosion products with relative dielectric permittivity of $\varepsilon_r=7$. In both studies, the thickness of the corroded products was in the order of 80 μm, a value which is well beyond the requirement for indoor applications.

McLaughlin *et al.* [17] proposed another type of UHF RFID device for corrosion monitoring. Indeed, instead of attaching the tag on the metal, these authors coated an UHF RFID tag with a resin containing a steel powder. Its thickness was higher than 25 μm. During corrosion of steel, clear changes of the reading rate of the tag were collected by the reader. The explanation of these results was based on the electromagnetic shielding of the antenna by the coating, in particular due the high magnetic permeability of corrosion products with respect to steel. The same approach based on electromagnetic shielding was used recently by considering steel foils of thickness higher than 50 μm [18].

Despite the interest of all these methods, the application of the RFID technology to the detection of loss of low thicknesses of metal such as copper remains an important issue, in particularly for indoor applications. The present study aims therefore at investigating the effects initiated by the coupling between an RFID antenna and a low thickness metallic layer exposed to a corrosive atmosphere. In the first part of the manuscript, electromagnetic simulations focus on such coupling by considering a dipole antenna which is traditionally used in the RFID technology. The achieved results will serve as a basis to explain the variation of the returned signal from a sensitive commercial RFID tag. As it will be highlighted experimentally, the method is suitable to detect low loss of metal in the range of tens of nanometers and should be considered in the future as a promising technique for indoor applications.

## 2. Material and methods

To investigate the $S_{11}$ reflection parameter of a dipole antenna coupled to a sensitive thin metallic element, electromagnetic simulations were realized using the software HFSS (High Frequency Structural Simulator) from ANSYS. It utilizes tetrahedral mesh elements to determine a solution to a given electromagnetic problem. Simulation sweeps were realized between 0.5 and 2 GHz, with an adaptive meshing at the frequency of interest: 868 MHz. At a later stage, to simulate the interrogation of a tag by a reader, the power received by an RFID the dipole was studied in front of a patch antenna excited by a wave port (input impedance = 50 Ω), playing the role of an RFID reader.

Experimental validation of the sensor was made by exposing a commercial RFID tag coupled with a thin sensitive layer of copper in a climatic chamber (Vötsh, VCC 0060) at a relative humidity of 90% and with a temperature ranging from 20°C to 35°C. Results were compared with those provided by electrical corrosion sensors (AirCorr™). The commercial RFID tag ALN-9654 "G" from Alien technologies with a Higgs™-3 IC chip was selected. The dimensions of the antenna printed on a substrate of 50 μm height are width = 93 mm, length = 19 mm, height = 10μm. The input impedance of the chip is $Z_{chip} = 28.6 - j.204$ Ω. The sensitive tag was interrogated in the climatic chamber by using a reader Impinj Speedway Revolution R420 operating with a 31 dBm power. A commercial patch antenna (S8658WPR) from LAIRD connectivity was connected to the reader and placed in the climatic chamber for the interrogation.

## 3. Results and discussion

### 3.1. Principle of the sensing property

An RFID tag is composed of an antenna and an RFID chip, which contains an identification number. It is energized by a radio frequency wave coming from an antenna/Reader and then returns back its identification code. The originality of this technology is to use battery-less passive tag on flexible substrate which displays thus very low visual nuisance and cost. Due to the autonomous property of tags, the reading distance *i.e.* the distance between the tag and the antenna/Reader is typically in the range of few meters. This value can be optimized by carefully designing antenna of both the reader and tag with respect to the environment. Within this technology, the backscattered power by the tag and collected by the reader can be obtained via the Friis equation :

$$P = P_R \tau \frac{R_{ant}}{R_{chip}} \left(\frac{G_R G_{ant}}{D^2}\right)^2 \left(\frac{\lambda}{4\pi}\right)^4 \qquad (1)$$

Where $P_R$ and $G_R$ are the power delivered by the reader and the gain of the reader's antenna, respectively. D is the reader to tag distance and λ the wavelength of the electromagnetic wave. The power P collected by the reader depends also on the gain $G_{ant}$ of the tag antenna and τ, the transmission coefficient factor which accounts for the mismatch between impedance of the antenna ($Z_{ant}$) and the impedance of the chip ($Z_{chip}$) [19]. The real part of the impedance $R_i$ (i=ant, chip) appears also on the relation. Note that the effect associated with different polarization orientation between the two antennas is not taken into account.

Most of commercial UHF RFID tags are composed of dipole antennas. These latter are selected since, with meandering techniques, it is possible to reduce their sizes and to ensure an impedance matching with the chip. Placing a dipole in a vicinity of a metallic object decreases the amplitude of electric field and consequently affects strongly its radiation pattern, efficiency and impedance ($G_{ant}$, $Z_{ant}$, τ in relation 1). This prevents the use of dipole attached to metallic object. This behavior is explained by the generation of a current with opposite phase, which results in the reduction of the radiation in the far field. However, to the best of our knowledge, the influence of a thin metallic layer with a thickness lower than 1 μm on the dipole antenna efficiency was not studied in the past. This point is of fundamental importance for the development of a sensitive RFID tag based on the coupling between the antenna and such thin metallic layer exposed to the environment. We therefore focus on this point by performing electromagnetic simulations. The sketch of the proposed device is depicted in Fig. 1. A dipole antenna is produced on a substrate characterized by dielectric properties ε'$_r$=2.1 , tanδ=0.001 and a thickness of 0.8 mm. This dipole is connected to a lumped port with a characteristic impedance of 73 Ω. The main parameters associated with the proposed dipole antenna resonating at 868 MHz without any sensitive element device are given in Fig. 1(b).

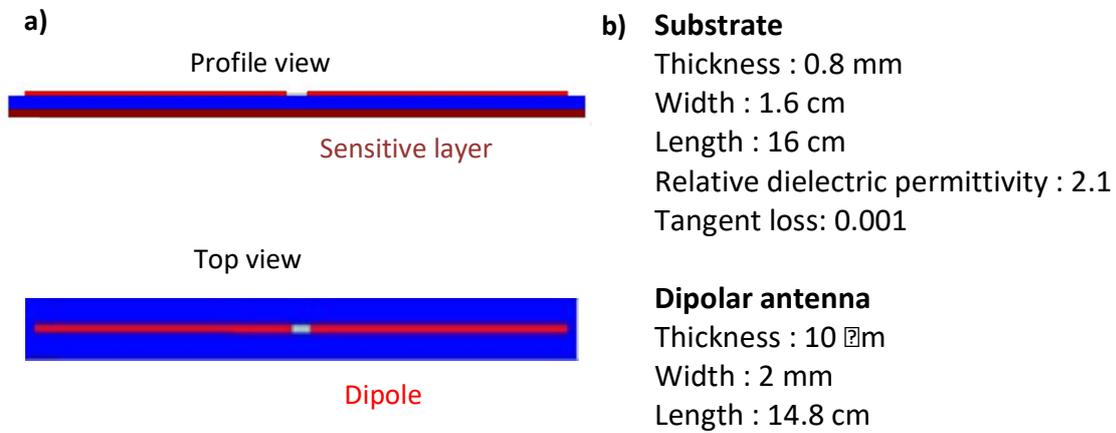

**Fig. 1.** (a) sketch of the coupling between the dipole antenna and the metallic layer; (b) dimensions of the dipole antenna.

As shown in Fig. 1(a), a sensitive layer to environmental corrosiveness is placed on the backside of the antenna. It is characterized by its thickness ($h_s$) and electrical conductivity ($\sigma$). The electrical conductivity $\sigma$ is chosen to be $10^7$ S.m$^{-1}$, a value close to the one of copper and silver which are used for IC class determination. To simulate the degradation of the sensitive element, we consider the corrosion process as uniform and explained by a homogenous loss of metal over the whole surface. In this case, the corrosion rate is simply defined by the loss of metal thickness divided by the time of exposure. HFFS simulations were consequently done for several thicknesses $h_s$ which correspond to different times of exposure during the corrosion of the film. Fig. 2(a-d) present by red lines the $S_{11}$ reflection parameter as function of the frequency for a thickness $h_s$ ranging from 10 μm to 10 nm. The results are compared with the data in black obtained for an insulating material ($\sigma=10^{-8}$ S/m) which simulates the final step of the corrosion process with the presence of insulating corrosion products on the backside of the sensor. As observed in Fig. 2(a-d), the dipole covered with this insulating layer exhibits a $S_{11}$ reflection parameter characteristic of the same dipole operating in air without any additional layer. The operating frequency, here 868 MHz, is obtained by considering a length of the dipole of 14.8 cm. As shown in Fig. 2(a), the presence of metallic layer with a thickness of 10 μm suppresses totally the resonant behavior. Fig. 2(b) displays the case of a layer of 1 μm thickness. The same behavior is observed even if this thickness is lower than the skin depth. In contrast, as seen in figures Fig. 2(c,d), a further decrease of $h_s$ to 100 nm and 10 nm leads to the appearance of the resonances. For a metal thickness of 10 nm, it is clearly identified with a minimal $S_{11}$ value of less than -10 dB which ensures a radiation efficiency and a resonant frequency of about 790 MHz. In these two cases, interrogating the dipole antenna might be feasible even if its efficiency remains mitigated with respect to the initial dipole antenna.

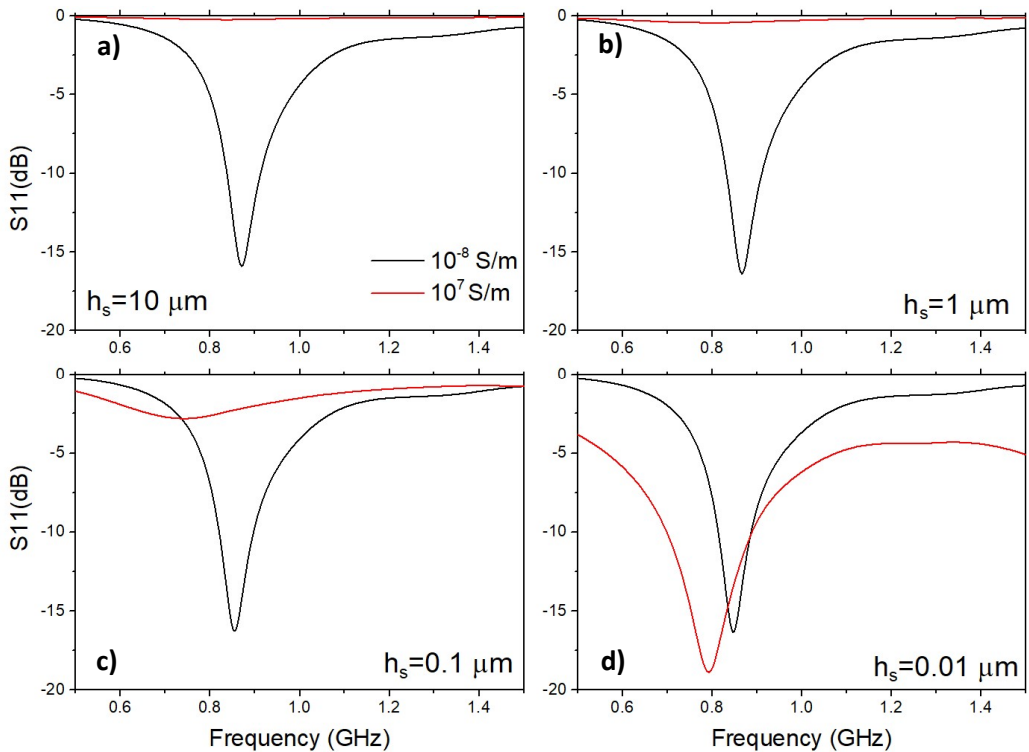

**Fig. 2.** Variations of the $S_{11}$ reflection parameter for different thicknesses $h_s$ of the sensitive layer: (a) $h_s$ = 10 μm, (b) $h_s$ = 1μm, (c) $h_s$ = 100 nm and (d) $h_s$ = 10 nm. The curves in red ($\sigma = 10^7$ S/m) represent the antenna covered by the metallic strip. The curves in black ($\sigma = 10^{-8}$ S/m) are associated to the final state with insulating corrosion products on the antenna.

To further get some insights on the main parameters which govern the proposed method and to compare the achieved results with those provided by electrical resistance sensors, it is interesting to consider the influence of the variation of electrical resistance R on the results. It is well known that, due to the relation between R, the thickness $h_s$, and the electrical conductivity ($\sigma$) of the metallic film, identical variations of R can be obtained by changing either $h_s$ or $\sigma$ by the same amount. Electromagnetic simulations were therefore made by considering a decrease the electrical conductivity ($\sigma$) of the metal from $\sigma = 10^8$ S/m to $10^5$ S/m for the selected values of $h_s$. Fig. 3(a,b) highlight the formation of the resonance when the conductivity is decreasing from $10^7$ S/m. A similar behavior was reported for a thickness decrease, indicating that the electrical resistance R of the sensitive layer is probably the main parameter which governs the antenna's efficiency changes. This result is confirmed by observing very similar $S_{11}$ reflection curves associated with one value of R but with different ($h_s$, $\sigma$) couples: ($10^8$ S/m, 10 nm); ($10^7$ S/m, 100 nm), ($10^6$ S/m, 1 µm), ($10^5$ S/m, 10 µm). In the same way, the $S_{11}$ reflection parameter associated with a sensitive layer of $h_s = 10$ nm thickness of metal ($\sigma = 10^7$ S/m) is very close to those found for ($10^6$ S/m, 100 nm); ($10^5$ S/m, 1 µm). From these results, the variation of the antenna property is thus strongly correlated to the variation of the electrical resistance of the sensitive layer as it is made for ER sensors.

The aim being to consider the power collected by the reader from the signal emitted by the modified tag, the influence of the variation of $h_s$ and $\sigma$ on the transmission of electromagnetic waves from this dipole antenna to a patch antenna localized at a distance of 15 cm was simulated. Fig. 4 displays the $S_{21}$ transmission parameter for thicknesses of 10 and 100 nm of metal ($\sigma = 10^7$ S/m) and for different values of the electrical conductivity which simulate the thickness decrease. The black curve corresponds to the dipole antenna with an insulating layer associated with the final corroded products. It is considered here as the reference.

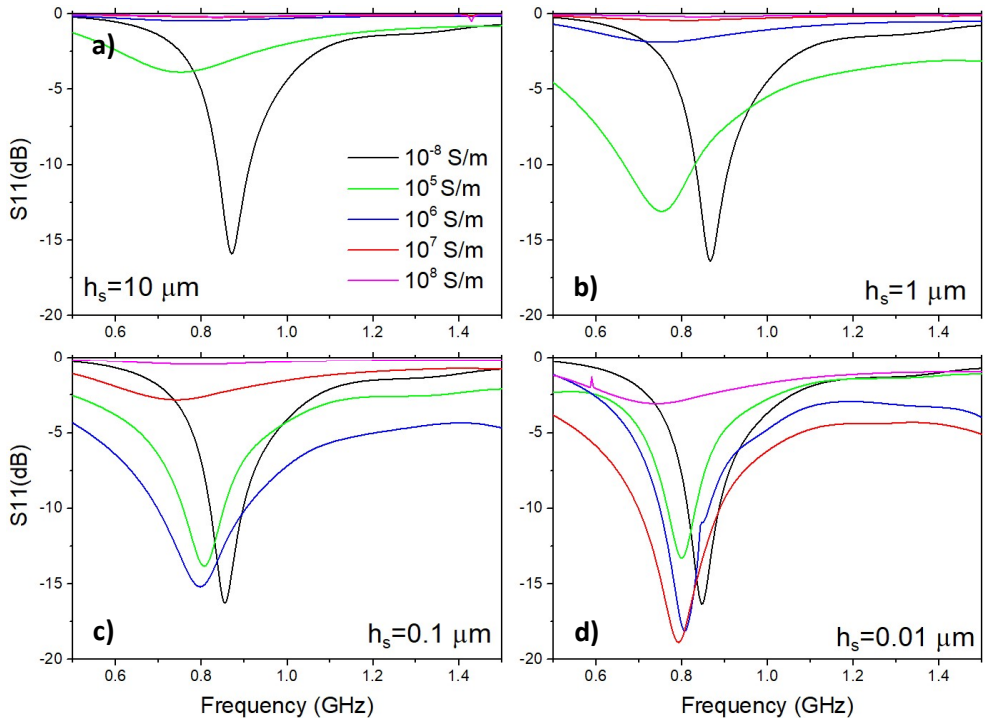

**Fig. 3.** Variations of the $S_{11}$ reflection parameter for different electrical conductivity $\sigma$ and thicknesses $h_s$ of the sensitive layer: (a) $h_s = 10$ μm, (b) $h_s = 1$ μm, (c) $h_s = 100$ nm and (d) $h_s = 10$ nm. The curves in red ($\sigma = 10^7$ S/m) represent the antenna covered by the metallic strip. The curves in black ($\sigma = 10^{-8}$ S/m) are associated to the final state with insulating corrosion products on the antenna.

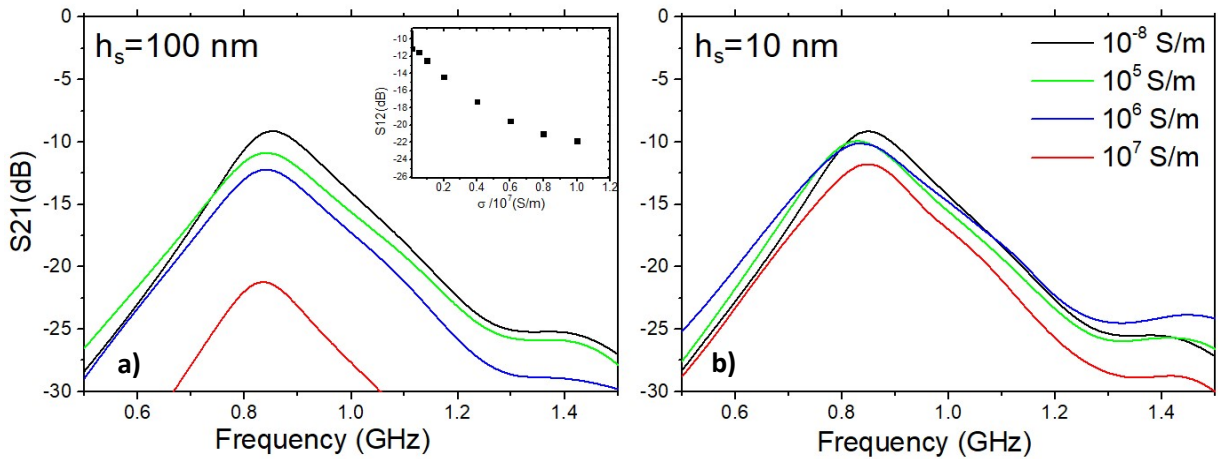

**Fig. 4.** Variation of the $S_{21}$ transmission parameter for several values of $\sigma$ for (a) $h_s = 100$ and (b) $h_s = 10$ nm. The curves in red ($\sigma = 10^7$ S/m) represent the antenna covered by the metallic strip. The curves in black ($\sigma = 10^{-8}$ S/m) are associated to the final state with insulating corrosion products on the antenna. Inset of (a): variation of the $S_{21}$ transmission parameter at 868 MHz for several values of $\sigma$.

In Fig. 4(a), as depicted by the red line, when considering a metallic sensitive layer of 100 nm thickness, the loss in transmission between the dipole and the patch antennas at 868 MHz which is the operating frequency is about -22 dB. A loss of efficiency of 12 dB with respect to the reference dipole (in black) is then obtained. The reduction of the electrical conductivity of the sensitive strip by a factor 10 which is similar to a decrease of its thickness to 10 nm, leads to an increase the $S_{21}$ level to -12.5 dB, a value close to that exhibited by the reference antenna. The inset of Fig. 4(a) displays the $S_{21}$ parameter measured at 868 MHz as function of the conductivity of the layer. The variation from -22 dB to -12.5 associated with the decrease of σ from $10^7$ to $0.1\ 10^7$ S/m is highlighted. As seen, it is almost linear from $0.8\ 10^7$ to $0.1\ 10^7$ S/m but deviates from a straight line for higher values of the conductivity. Fig. 4(b) displays the results for a metal thickness of 10 nm: an $S_{21}$ level of -12.5 dB, in very good agreement with the data of Fig. 4(a) is observed. This further proves that the electrical resistance governs the antenna efficiency variation. From these electromagnetic simulations, it appears clearly that monitoring the loss of thickness in the range 10/100 nm via the measurements of the power collected by the reader should be feasible. Indeed, a variation of 9.5 dB of the $S_{21}$ level is correlated to a thickness changing from 100 to 10 nm leading to a sensitivity of about 1dB/nm. The sensitive layer is then very similar to that used in ER sensors. Indeed, in this case, layers with initial thicknesses of few hundreds of nanometers are integrated to ensure a good accuracy and lifetime.

## 3.2. Experimental validation

To validate experimentally the monitoring of the loss of metal in the range of 100/10 nm by the proposed device, a commercial UHF RFID tag was considered. The Corrosiveness Tag (CTag) is presented in Fig. 5. The first side of the CTag displays the conventional antenna and RFID chip on a flexible substrate (Fig 5(a)). The other side (Fig. 5(b)) is composed of a thin film of copper layer with a thickness of 30 nm elaborated by magnetron sputtering. This method of elaboration is very reproductive in terms of thickness and produces samples with very low roughness. A thickness of 30 nm was chosen since a loss of this amount of copper per year is close to the threshold between the IC2 (low corrosivity) and IC3 (middle corrosivity) classes. The thin copper layer was first elaborated on a polycarbonate film with a thickness of 0.1 mm. The adhesive commercial tag was then placed on the polycarbonate substrate. The proposed Ctag differs from device used for simulations. Indeed, as seen in Fig. 5(a), the shape of the dipole antenna is strongly modified to match the impedance of the chip. Despite this difference, the simulated result will serve as basis to describe the experimental results. During the exposure test in the climatic chamber, a second tag identical to CTag but without the copper layer is considered. Due to the absence of copper, it is considered as the reference tag.

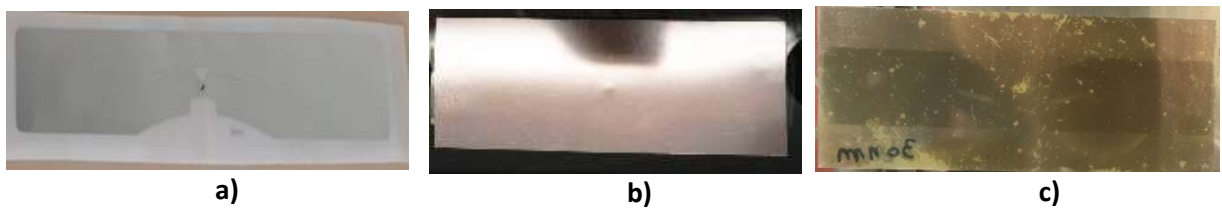

**Fig. 5.** (a) side of CTag containing the RFID tag and antenna; (b) second side before exposure; (c) second side after exposure. The dimension of the tag antenna is 9 cm x 2cm.

To test the atmospheric corrosion of the sensitive layer on the RFID communication, the CTag was placed in a climatic chamber (90%RH, variable temperature: 20°C, 35°C). Two ER sensor (Aircorr$^{TM}$) equipped with a sensitive layer of copper (thickness = 250 nm) and a reference uncovered tag were also used. The RFID tags were interrogated by an Impinj 420 Reader operating at a power of 31 dBm. The reader antenna was placed in the climatic chamber at a distance of about 15 cm of the tags. The results are depicted in Fig. 6.

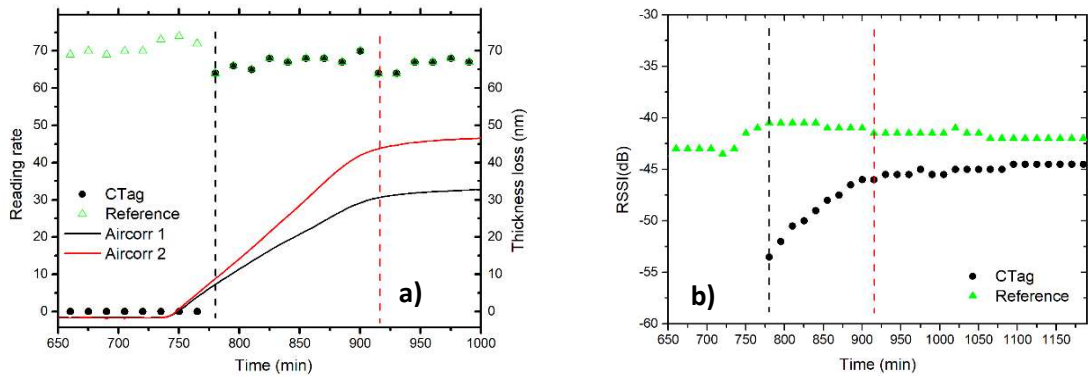

**Fig. 6.** (a) Reading rate of the CTag and reference tags compared to the data collected by two ER sensors during exposure. (b) RSSI factor of the CTag and reference tags.

Fig. 6(a) displays the loss of metallic copper as function of the exposure time obtained by the two electrical resistance sensors. During the measurements, the variation of the electrical resistance of the strips was measured and converted to thickness loss. At low exposure time, with a temperature of 20°C, almost no variation of the thickness is observed due to the low corrosivity of the environment. At about 750 min exposure time, when increasing the temperature to 35°C, a clear loss of metal thickness appears. It depends almost linearly of the exposure time giving a constant corrosion rate *i.e.* the thickness loss per time unit. After 900 minutes, a reduction of the corrosion rate leading to a parabolic shape is shown, which is typical for copper corrosion under atmospheric conditions [20]. As seen the curves provided by the two sensors are not perfectly identical. The deviation of extreme values from the mean value is 16%, a value which not very far from that achieved by Kouril *et al.* [2]. This variability should be compared with other techniques used to measure the loss of thickness, namely coupons. Within the discussed range of tens of nanometers the variability of this method exceeds largely the presented results.

For the CTag, two parameters can be collected by the RFID reader. The reading rate is the number of time that the tag responds to the reader over a fixed period of interrogation time. Fig. 6(a) presents the reading rate of the CTag and reference tag during the exposure. At the beginning of the test, no response (reading rate = 0) of the CTag is observed due to the presence of the metallic copper layer of 30 nm thickness In contrast, the reference tag is responding. After 775 minutes of exposure time, a transition occurs and the reader is able to see the CTag. At this time represented by black dashed line in the figure, the CTag received a sufficient amount of power to operate. The reading rate of the CTag is then identical to that found for the reference tag. The second parameter of interest is the RSSI factor (Received Signal Strength Indicator). It is a measurement of the power received from the returned signal from an RFID tag when interrogated by the reader. It therefore refers to the backscattered power P of equation (1) and to the $S_{21}$ parameter investigated above. Fig. 6(b) displays the RSSI factor as function of the exposure time for both the CTag and reference tags. As discussed previously, the first response of the CTag occurs at about 775 min when it starts to be energized. A RSSI value of -53.5 dB is then found. A further decrease of the metal thickness due to corrosion leads to an increase of the RSSI, in very good agreement with our theoretical expectations. After about 920 minutes of exposure time, as indicated by red dashed line in the figure, the RSSI value becomes constant at -44.5 dB. During, the exposure time, the reference tag responds to the reader with an almost constant value of the RSSI (-42 dB). In the final state, this value differs by 2.5 dB from the RSSI of the CTag. This is explained by different localizations of both tags in the climatic chamber, a complex location where of reflection of EM waves occurs.

These results can be compared with those obtained by ER sensors. From the data presented in Fig. 6(a), the first transition in black occurs at 775 minutes when a thickness of about 7 nm of copper is lost by the sensitive layer. With a thickness of 23 nm copper, the tag is then energized and is able to respond for the first time to the reader. The second transition, in red, corresponds to the loss of the initial copper thickness of 30 nm. Fig. 5(c) which displays the backside of CTag after exposure. As seen, the initial copper layer is strongly corroded leading to the presence of a very thin layer of corroded products. The antenna localized at the other side can then be observed by transparency. The variation of the RSSI from -53.5 dB to -

44.5 dB is associated with these two transitions. The sensitivity of the method is therefore equal to 0.4 dB/nm and authorizes to probe the thresholds from IC1 to IC2 (6 nm/year) and from IC2 to IC3 (22 nm/year). This confirms the interest of the method for low corrosiveness classes determination. Note that the sensitivity is definitely higher than the one achieved by electromagnetic simulations where a thinning from 100 nm to 10 nm resulting to about 10 dB (0.1 dB/nm). This is explained by a different shape of the antenna and impedance of the input port with respect to simulations.

**5. Conclusion**s

As a conclusion, we propose a new method for atmospheric corrosion monitoring for indoor applications which enables to follow the loss of thickness of metals at the nanometer scale. It is based on the variation of the signal collected by an RFID reader from a device constituted of an RFID chip coupled with a thin metallic layer exposed to the environment. As shown, the sensitivity of the method is ensured by the increase of the electrical resistance of the sensitive layer due to its corrosion. The experimental results achieved with a copper layer of 30 nm thickness demonstrate indeed the ability of the method with a sensitivity of 0.4 dB/nm. Due the low cost and low visual nuisance of the proposed tag, the method should therefore be considered as a promising technique for the development of new applications concerning corrosion monitoring.

**Acknowledgements**

This work received funding from the European Union's Horizon H2020 research and innovation program under grant agreement 814596 (Preventive solutions for Sensitive Materials of Cultural Heritage – SENSMAT).

**References**
[1] ISO 11844-1, Corrosion of metals and alloys – classification of low corrosivity of indoor atmospheres – Part 1: determination and estimation of indoor corrosivity, International Organization of Standardization, Genève, Suisse, 2006.

[2] M. Kouril, T. Prosek, B. Scheffel and F. Dubois, High sensitivity electrical resistance sensors for indoor corrosion monitoring, Corrosion Engineering Science and Technology 48 (2013) 282–287.

[3] M. Kouril, T. Prosek, B. Scheffel, Y. Degres, Corrosion monitoring in archives by the electrical resistance technique, Journal of Cultural Heritage 15 (2014) 99–103.

[4] T. Prosek, M. Taube, F. Dubois, D. Thierry, Application of automated electrical resistance sensors for measurement of corrosion rate of copper, bronze and iron in model indoor atmospheres containing short-chain volatile carboxylic acids, Corrosion Science 87 (2014) 376–382.

[5] J. Zhang, G. Yun Tian, A. M. J. Marindra, A. Imam Sunny and A. Bo Zhao, A Review of Passive RFID Tag Antenna-Based Sensors and Systems for Structural Health Monitoring Applications, Sensors 17 (2017) 265.

[6] C. Occhiuzzi, S. Caizzone, G. Marrocco, Passive UHF RFID Antennas for Sensing Applications: Principles, Methods and Classifications, IEEE Antennas and Propagation Magazine 55 (2013) 14-34.

[7] K. Mc Gee, P. Anandarajah and D. Collins, A Review of Chipless Remote Sensing Solutions Based on RFID Technology, Sensors 19 (2019) 4829.

[8] C. Herrojo, F. Paredes, J. Mata-Contreras and F. Martín, Chipless-RFID: A Review and Recent Developments, Sensors 19 (2019) 3385.

[9] V. Mulloni and M. Donelli ,yyjChipless RFID Sensors for the Internet of Things: Challenges and Opportunities, Sensors 20 (2020) 2135.

[10] M. Forouzandeh and N.C. Karmakar, Chipless RFID tags and sensors: a review on time-domain techniques, Wireless Power Transfer 2 (2015) 62–77.

[11] M. Yasri, B. Lescop, E. Diler, F. Gallée, D. Thierry and S. Rioual, Fundamental basis of electromagnetic wave propagation in a zinc microstrip lines during its corrosion, Sens. Actuators B : Chem. 223 (2016) 352-358.

[12] R. Khalifeh, M. Yasri, B. Lescop, F. Gallée, E. Diler, D. Thierry, S. Rioual, Development of wireless and passive corrosion sensors for material degradation monitoring in coastal zones and immersed environment, IEEE J. Ocean. Eng. 99 (2016) 776-782.

[13] M. Yasri, B. Lescop, E. Diler, F. Gallée, D. Thierry and S. Rioual, Monitoring uniform and localised corrosion by a radiofrequency sensing method, Sens. Actuators B : Chem. 257 (2018) 988-992.

[14] J. Rammal, F. Salameh, O. Tantot, N. Delhote, S. Verdeyme, S. Rioual, F. Gallé, B. Lescop, A Microwave sensor for zinc corrosion detection, J. Appl. Phys. 122 (2017) 114501.

[15] J. Zhang and G. Yun Tian, UHF RFID Tag Antenna-Based Sensing for Corrosion Detection & Characterization Using Principal Component Analysis, IEEE Trans. Antennas Propag. 64 (2016) 4405–4414.

[16] S. Soodmand, A. Zhao, G. Yun Tian, UHF RFID system for wirelessly detection of corrosion based on resonance frequency shift in forward interrogation power, IET Microw. Antennas Propag. 12 (2018) 1877-1884.

[17] S. R. McLaughlin , Y. He and J. Lo, Development of a Wireless Corrosion Monitoring Sensor for Land Vehicles Development of a Wireless Corrosion Monitoring Sensor for Land Vehicles, Conference: 2015 Department of Defense - Allied Nations Technical Corrosion Conference at Pittsburgh, PA

[18] Y. He, Wireless Corrosion Monitoring Sensors Based on Electromagnetic Interference Shielding of RFID Transponders, Corrosion 76 (2020) 411-423.

[19] C. Occhiuzzi, S. Cippitelli, and G. Marrocco, Modeling, Design and Experimentation of Wearable RFID Sensor Tag, IET Microw. Antennas Propag. 58 (2010) 58.

[20] C. Leygraf and T. Graedel, Atmospheric corrosion, 38, 103; 2000, New York, Wiley Interscience.